\documentclass[aps,prd,preprint,groupedaddress]{revtex4-1}
\usepackage{amssymb}
\usepackage{graphicx}
\begin{document}


\title{Distinguishing between $\Lambda$CDM and modified gravity \\
with future observations of cosmic growth rate}


\author{Koichi Hirano}
\email[E-mail: ]{k\_hirano@tsuru.ac.jp}
\affiliation{Department of Primary Education, Tsuru University, 3-8-1, Tahara, Tsuru, Yamanashi 402-8555, Japan}



\begin{abstract}
A probability of distinguishing between $\Lambda$CDM model and modified gravity is studied by using the future observations for the growth rate of cosmic structure (Euclid redshift survey). Adopting extended DGP model, Kinetic Gravity Braiding model, and Galileon model as modified gravity, we compare predicted cosmic growth rate by models to the mock observational data. The growth rate $f\sigma_8$ in the original DGP model is suppressed compared with the $\Lambda$CDM case, for the same value of the current density parameter of matter $\Omega_{m,0}$, because of the suppression of effective gravitational constant. In case of the kinetic gravity braiding model and the Galileon model, the growth rate $f\sigma_8$ is enhanced compared with the $\Lambda$CDM case, for the same value of $\Omega_{m,0}$, because of the enhancement of effective gravitational constant. For future observational data of the cosmic growth rate (Euclid), compatible value of $\Omega_{m,0}$ are different by models, furthermore $\Omega_{m,0}$ can be stringently constrained. Thus, we find the $\Lambda$CDM model is distinguishable from modified gravity by combining the growth rate data of the Euclid with other observations.
\end{abstract}


\maketitle

\section{INTRODUCTION}

Cosmological observations, including type Ia supernovae (SNIa) \cite{rie1998,per1999}, the cosmic microwave background (CMB) anisotropies, and baryon acoustic oscillations (BAO) indicate that the universe is undergoing an accelerated phase of expansion. This late-time acceleration is one of the biggest mysteries in current cosmology. The standard explanation is that it is caused by the dark energy \cite{rat1988,cal2002,kom2006,kom2005}. This would mean that the universe is mostly filled with an unknown component. The cosmological constant is the standard candidate of dark energy. To explain the current acceleration of the universe, the cosmological constant must have an considerably small value. However, its value cannot be explained by particle physics and it is affected by fine-tuning problems and the coincidence problem.

An alternative explanation for the current accelerated expansion of the universe is to extend general relativity to a more general theory of gravity at long scale. Several modified gravity approaches have been proposed  such as $f(R)$ gravity \cite{fel2010}, scalar-tensor theories \cite{hir2010c,hir2008a,har2008}, and the Dvali--Gabadazde--Porrati (DGP) braneworld model \cite{dva2000,hir2010a,hir2011a}.

Further, as an alternative to general relativity, Galileon gravity models have been proposed \cite{nic2009,cho2009,sil2009,kob2010a,kob2010b,gan2010,fel2010a,fel2010b,def2010}. These models are built by introducing a scalar field with a self-interaction whose Lagrangian is invariant under Galileon symmetry $\partial_\mu\phi\rightarrow\partial_\mu\phi+b_\mu$, which keeps the equation of motion as a second-order differential equation. This prevents the theory from exhibiting a new degree of freedom, and perturbation of the theory does not raise ghost or instability problems. The simplest term of the self-interaction is $\Box\phi(\nabla\phi)^2$, which appears in the four-dimensional effective theory of the DGP model. The self-interaction term $\Box\phi(\nabla\phi)^2$ induces decoupling of the Galileon field $\phi$ from gravity at small scales by the Vainshtein mechanism \cite{vai1972}. This allows the Galileon theory to recover general relativity at scales around the high density region, which is consistent with solar system experiments.

Galileon theory has been covariantized and studied in curved backgrounds \cite{def2009a,def2009b}. It has been shown that Galileon symmetry cannot be preserved once the theory is covariantized, however it is possible to maintain the equation of motion as a second-order differential equation, that is, free from ghost-like instabilities. Galileon gravity induce self-accelerated expansion of the late-time universe. Thus, inflation models inspired by Galileon theory have been proposed \cite{kob2010c,miz2010,kam2010}. In Ref. \cite{nes2010}, the parameters of the generalized Galileon cosmology were constrained from observational data of supernovae Ia (SNIa), the cosmic microwave background (CMB) and baryon acoustic oscillations (BAO). The evolution of matter density perturbations for Galileon models have also been studied \cite{kim2011,fel2010c,sil2009,kob2010a,kob2010b}.

Almost forty years ago, Horndeski derived the action of the most general scalar-tensor theories with second-order equations of motion \cite{horndeski1974}. This theory recently received much attention as an extension of covariant Galileons \cite{nic2009,def2009a,def2009b,charmousis2012}. One can show that the four-dimensional action of generalized Galileons derived by Deffayet et al. \cite{deffayet2011} is equivalent to the Horndeski's action under a field redefinition \cite{kobayashi2011}. Since this theory contains all modified gravity models and the single-field inflation models with one scalar degree of freedom as specific cases, it is very important in cosmology and hence considerable attention has been paid recently to various aspects of Horndeski's theory.

In this paper, a probability of distinguishing between $\Lambda$CDM model and modified gravity is studied by using the future observations for the growth rate of cosmic structure (Euclid redshift survey \cite{amendola2012}). we computed the growth rate of matter density perturbations in modified gravity and compared it with mock observational data.
Whereas the background expansion history in modified gravity is almost identical to that of dark energy models, the evolution of matter density perturbations of modified gravity is different from that of dark energy models.
Thus, it is important to study the growth history of perturbations to distinguish modified gravity from models based on the cosmological constant or dark energy.

Using the past observations of the growth rate of matter density perturbations, modified gravity have been studied. \cite{okada2013}. On the other hand, we focus on the future observations of the growth rate by Euclid. We adopt extended DGP model \cite{dva2003}, Kinetic Gravity Braiding model \cite{kim2011}, and Galileon model \cite{sil2009,kob2010a} as modified gravity models. Kinetic Gravity Braiding model and Galileon model are specific aspects of Horndeski's theory.

This paper is organized as follows. In the next section, we present the background evolution and the effective gravitational constant in modified gravity models.
In Section \ref{growthrate}, we describe the theoretical computations and the mock observational data of the growth rate of matter density perturbations. In Section \ref{comparison}, we study a probability of distinguishing between $\Lambda$CDM model and modified gravity, by comparison of predicted cosmic growth rate by models to the mock observational data. Finally, conclusions are given in Section \ref{conclusion}.

\section{MODIFIED GRAVITY MODELS}

\subsection{extended DGP model}


In the DGP model\cite{dva2000}, it is assumed that we live on a 4D brane embedded in a 5D Minkowski bulk. Matter is trapped on the 4D brane and only gravity experiences the 5D Minkowski bulk.

The action is
\begin{equation}
S = \frac{1}{16\pi}M_{(5)}^3\int_{bulk}{d^5x\sqrt{-g_{(5)}}R_{(5)}} + \frac{1}{16\pi}M_{(4)}^2\int_{brane}{d^4x\sqrt{-g_{(4)}}(R_{(4)}+L_m)},
\end{equation}
where the subscripts (4) and (5) denote quantities on the 4D brane and in the 5D bulk, respectively. $M_{(5)}$ ($M_{(4)}$) represents the 5D (4D) Planck mass, and $L_m$ is the matter Lagrangian confined on the brane. The transition from 4D to 5D gravity is governed by a crossover scale $r_c$.
\begin{equation}
r_c = \frac{M_{(4)}^2}{2M_{(5)}^3}.
\end{equation}
At scales larger than $r_c$, gravity appears in 5D. At scales smaller than $r_c$, gravity is effectively bound to the brane and 4D Newtonian dynamics is recovered to a good approximation. $r_c$ is a parameter in this model, which has a unit of length \cite{koyama2006}.

Under spatial homogeneity and isotropy, a Friedmann-like equation is obtained on the brane \cite{def2001,def2002}:
\begin{equation}
H^2 = \frac{8\pi G}{3}\rho+\epsilon\frac{H}{r_c} \label{dgp_fri},
\end{equation}
where $\rho$ is the total cosmic fluid energy density on the brane. the DGP model have the two branches ($\epsilon = \pm 1$). The solution with $\epsilon = +1$ is the self-accelerating branch. In this branch, the expansion of the universe accelerates without dark energy, because the Hubble parameter approaches a constant $H = 1/r_c$, at late times. On the other hand, $\epsilon = -1$ is the normal branch. This branch cannot undergo acceleration without an dark energy component. Hence, in what follows, we consider the self-accelerating branch only ($\epsilon = +1$).




The original DGP model, however, is plagued by the ghost problem \cite{koy2007} and is incompatible with cosmological observations \cite{xia2009}.


Dvali and Turner \cite{dva2003} phenomenologically extended the Friedmann-like equation of the DGP model (Eq. (\ref{dgp_fri})). This model interpolates between the original DGP model and the $\Lambda$CDM model using an additional parameter $\alpha$. The modified Friedmann-like equation is
\begin{equation}
H^2 = \frac{8\pi G}{3}\rho+\frac{H^{\alpha}}{{r_c}^{2-\alpha}}. \label{dt_fri}
\end{equation}
For $\alpha = 1$, this is equivalent to the original DGP Friedmann-like equation, while $\alpha = 0$ leads to an expansion history identical to $\Lambda$CDM cosmology. This is important in order to distinguish the $\Lambda$CDM model from the original DGP model between $\alpha = 0$ and $1$.

In case of the extended DGP model, crossover scale $r_c$ can be expressed as follows:
\begin{equation}
r_c = (1-\Omega_{m,0})^{\frac{1}{\alpha-2}}H_0^{-1}.
\end{equation}
Thus independent parameters as cosmological model are $\alpha$ and the today's energy density parameter of matter $\Omega_{m,0}$. The effective gravitational constant of the extended DGP model is given so as to interpolate between $\Lambda$CDM and original DGP model.
\begin{equation}
\frac{G_{\rm eff}}{G} = 1+\frac{1}{3\beta},
\end{equation}
where
\begin{equation}
\beta \equiv 1-\frac{2(r_cH)^{2-\alpha}}{\alpha}\left[1+\frac{1}{3}\frac{(2-\alpha)\dot{H}}{H^2}\right].
\end{equation}
$G_{\rm eff}/G$ is the effective gravitational constant normalized to Newton's gravitational constant, and an overdot represents differentiation with respect to cosmic time $t$.

\subsection{Kinetic Gravity Braiding model}

Kinetic gravity braiding model \cite{kim2011} is proposed as an alternative to the dark energy model. One can say that kinetic gravity braiding model is specific aspects of Horndeski's theory \cite{horndeski1974}.

The most general four-dimensional scalar-tensor theories keeping the field equations of motion at second order are described by the Lagrangian \cite{horndeski1974,deffayet2011,charmousis2012,kobayashi2011,felice2011}
\begin{equation}
\mathcal{L} = \sum^5_{i=2}\mathcal{L}_i, \label{lag_horn}
\end{equation}
where
\begin{eqnarray}
\mathcal{L}_2 & = & K(\phi,X), \\
\mathcal{L}_3 & = & -G_3(\phi,X)\Box\phi,  \\
\mathcal{L}_4 & = & G_4(\phi,X)R+G_{4,X}[(\Box\phi)^2-(\nabla_\mu\nabla_\nu\phi)(\nabla^\mu\nabla^\nu\phi)], \\
\mathcal{L}_5 & = & G_5(\phi,X)G_{\mu\nu}(\nabla^\mu\nabla^\nu\phi)-\frac{1}{6}G_{5,X}[(\Box\phi)^3-3(\Box\phi)(\nabla_\mu\nabla_\nu\phi)(\nabla^\mu\nabla^\nu\phi)+2(\nabla^\mu\nabla_\alpha\phi)(\nabla^\alpha\nabla_\beta\phi)(\nabla^\beta\nabla_\mu\phi)]. \nonumber \label{lag_horn5} \\
\end{eqnarray}
Here $K$ and $G_i$ ($i = 3, 4, 5$) are functions of a scalar field $\phi$ and its kinetic energy $X = -\partial^\mu\phi\partial_\mu\phi/2$, with the partial derivatives $G_{i,X} \equiv \partial G_i/\partial X$. $R$ is the Ricci scalar, and $G_{\mu\nu}$ is the Einstein tensor. The above Lagrangian was first derived by Horndeski in a different form \cite{horndeski1974}. This Lagrangian (Eqs. \ref{lag_horn}-\ref{lag_horn5}) is equivalent to that derived by Horndeski \cite{kobayashi2011}. The total action is then given by
\begin{equation}
S = \int d^4x\sqrt{-g}(\mathcal{L}+\mathcal{L}_m),
\end{equation}
where $g$ represents a determinant of the metric $g_{\mu\nu}$, and $\mathcal{L}_m$ is the Lagrangian of non-relativistic matter.

Variation with respect to the metric produces the gravity equations, and variation with respect to the scalar field $\phi$ yields the equation of motion. Using the following notation $K\equiv K(\phi,X)$, $G\equiv G_3(\phi,X)$, $F\equiv\frac{2}{M_{\rm pl}^2}G_4(\phi,X)$, and assuming $G_5(\phi,X)=0$, for Friedmann--Robertson--Walker spacetime, the gravity equations give
\begin{equation}
3M_{\rm pl}^2FH^2=\rho_m+\rho_r-3M_{\rm pl}^2H\dot{F}-K+2XK_{,X}+6H\dot{\phi}XG_{,X}-2XG_{,\phi}, \label{freq1}
\end{equation}
\begin{equation}
-M_{\rm pl}^2F(3H^2+2\dot{H})=p_r+2M_{\rm pl}^2H\dot{F}+M_{\rm pl}^2\ddot{F}+K-2XG_{,X}\ddot{\phi}-2XG_{,\phi}, \label{freq2}
\end{equation}
and the equation of motion for the scalar field gives
\begin{eqnarray}
& & (K_{,X}+2XK_{,XX}+6H\dot{\phi}G_{,X}+6H\dot{\phi}XG_{,XX}-2XG_{,\phi X}-2G_{,\phi})\ddot{\phi} \nonumber \\
& & +(3HK_{,X}+\dot{\phi}K_{,\phi X}+9H^2\dot{\phi}G_{,X}+3\dot{H}\dot{\phi}G_{,X}+6HXG_{,\phi X}-6HG_{,\phi}-G_{,\phi\phi}\dot{\phi})\dot{\phi} \nonumber \\
& & -K_{,\phi}-6M_{\rm pl}^2H^2F_{,\phi}-3M_{\rm pl}^2\dot{H}F_{,\phi}=0, \label{eom_scalar}
\end{eqnarray}
here an overdot denotes differentiation with respect to cosmic time $t$ and $H=\dot{a}/a$ is the Hubble expansion rate. Note that we use the following partial derivative notation: $K_{,X}\equiv\partial K/\partial X$ and $K_{,XX}\equiv\partial^2K/\partial X^2$, and similarly for other variables.
$\rho_m$ and $\rho_r$ are the energy densities of matter and radiation, respectively, and $p_r$ is the pressure of the radiation.

Kinetic gravity braiding model \cite{kim2011} correspond to a case that the functions in the Horndeski's theory are given as follows:
\begin{eqnarray}
K(\phi,X) & = & -X, \\
G_3(\phi,X) & = & M_{\rm pl}\left(\frac{r_c^2}{M_{\rm pl}^2}X\right)^n, \label{kgb_n} \\
G_4(\phi,X) & = & \frac{M_{\rm pl}^2}{2}, \\
G_5(\phi,X) & = & 0.
\end{eqnarray}
$M_{\rm pl}$ is the reduced Planck mass related with Newton's gravitational constant by $M_{\rm pl} = 1/\sqrt{8\pi G}$. where
$r_c$ is called the crossover scale in the DGP model \cite{koyama2006}.
The kinetic braiding model we study is characterized by a parameter $n$ in Eq. (\ref{kgb_n}), which corresponds to the Deffayet's galileon cosmological model for $n = 1$ \cite{def2010}. The background expansion of the universe of the kinetic braiding model approaches to that of the $\Lambda$CDM model for $n$ equal to infinity. This is important in order to distinguish between the kinetic braiding model and the $\Lambda$CDM model when $n$ is finite.

In case of the kinetic braiding model, using the Hubble parameter as the present epoch $H_0$, the crossover scale $r_c$ is given by
\begin{equation}
r_c = \left(\frac{2^{n-1}}{3n}\right)^{1/2n}\left[\frac{1}{6(1-\Omega_{m,0}-\Omega_{r,0})}\right]^{(2n-1)/4n}H_0^{-1},
\end{equation}
where $\Omega_{r,0}$ is the density parameter of the radiation, at present. Thus independent parameters as cosmological model are $n$ and 
$\Omega_{m,0}$. The effective gravitational constant normalized to Newton's gravitational constant $G_{\rm eff}/G$ of the kinetic braiding model is given by.
\begin{equation}
\frac{G_{\rm eff}}{G} = \frac{2n+3n\Omega_m-\Omega_m}{\Omega_m(5n-\Omega_m)},
\end{equation}
where $\Omega_m$ is the matter energy density parameter defined as $\Omega_{m}=\rho_m/3M_{\rm pl}^2H^2$. Here, we used the attractor condition. Although the background evolution for large $n$ approaches the $\Lambda$CDM model, the growth history of matter density perturbations is different due to the time-dependent effective gravitational constant.

\subsection{Galileon model}


Galileon gravity model is proposed as an alternative to the dark energy model. One can say that the Galileon model studied in Refs. \cite{sil2009,kob2010a} is specific aspects of Horndeski's theory \cite{horndeski1974}.

The Galileon model \cite{sil2009,kob2010a} correspond to a case that the functions in the Lagrangian (equations (\ref{lag_horn})-(\ref{lag_horn5})) of the Horndeski's theory are given as follows:
\begin{eqnarray}
K(\phi,X) & = & 2\frac{\omega}{\phi}X, \label{scf2} \\
G_3(\phi,X) & = & 2\xi(\phi)X, \label{scf3} \\
G_4(\phi,X) & = & \phi, \label{scf4} \\
G_5(\phi,X) & = & 0 \label{scf5},
\end{eqnarray}
where $\omega$ is the Brans--Dicke parameter and $\xi(\phi)$ is a function of $\phi$.

In this case, the gravity equations (the Friedmann-like equations) (\ref{freq1}) and (\ref{freq2}) can be written in the following forms, respectively:
\begin{equation}
3H^2=\frac{1}{M_{\rm pl}^2}(\rho_m+\rho_r+\rho_\phi) \label{frefre1},
\end{equation}
\begin{equation}
-3H^2-2\dot{H}=\frac{1}{M_{\rm pl}^2}(p_r+p_\phi) \label{frefre2},
\end{equation}
where the effective dark energy density $\rho_\phi$ is defined as
\begin{equation}
\rho_\phi = 2\phi\left[-3H\frac{\dot{\phi}}{\phi}+\frac{\omega}{2}\left(\frac{\dot{\phi}}{\phi}\right)^2+\phi^2\xi(\phi)\left\{3H+\frac{\dot{\phi}}{\phi}\right\}\left(\frac{\dot{\phi}}{\phi}\right)^3\right]+3H^2\left(M_{\rm pl}^2-2\phi\right), \label{rho_phi}
\end{equation}
and the effective pressure of dark energy $p_\phi$ is
\begin{eqnarray}
p_\phi & = & 2\phi\left[\frac{\ddot{\phi}}{\phi}+2H\frac{\dot{\phi}}{\phi}+\frac{\omega}{2}\left(\frac{\dot{\phi}}{\phi}\right)^2-\phi^2\xi(\phi)\left\{\frac{\ddot{\phi}}{\phi}-\left(\frac{\dot{\phi}}{\phi}\right)^2\right\}\left(\frac{\dot{\phi}}{\phi}\right)^2\right] \nonumber \\
 & & -(3H^2+2\dot{H})\left(M_{\rm pl}^2-2\phi\right). \label{p_phi}
\end{eqnarray}
The equation of motion for the scalar field is given by equation (\ref{eom_scalar}).

For the numerical analysis,
we adopt a specific model in which
\begin{equation}
\xi(\phi)=\frac{r_c^2}{\phi^2}, \label{scf6}
\end{equation}
where $r_c$ is the crossover scale \cite{cho2009}. This Galileon model is the Brans-Dicke theory extended by adding the self-interaction term: $\xi(\phi)(\nabla\phi)^2\Box\phi$. Thus, $\omega$ of this model is not exactly the same as the original Brans-Dicke parameter.
The evolution of matter density perturbations of this model has been computed in Refs. \cite{sil2009,kob2010a}.

At early times, to recover general relativity, we set the initial condition $\phi\simeq M_{\rm pl}^2/2$. This reduces the Friedmann equations (Eqs. (\ref{frefre1}) and (\ref{frefre2})) to usual forms: $3H^2 = {(\rho_m+\rho_r)}/M_{\rm pl}^2$ and $-3H^2 - 2\dot{H}={p_r}/M_{\rm pl}^2$. This is the cosmological version of the Vainshtein effect \cite{vai1972}, the method by which general relativity is recovered below a certain scale. At present, to induce the cosmic acceleration, the value of $r_c$ must be fine-tuned.



Since the energy density parameter of the matter at present in this model is defined as $\Omega_{m,0}=\rho_{m,0}/3H_0^2\phi_0$, in the numerical analysis, the value of $r_c$ be fine-tuned so that $\Omega_{m,0}$ becomes an assumed value. Thus independent parameters as cosmological model are $\omega$ and $\Omega_{m,0}$. For the Galileon model specified by Eqs. (\ref{scf2})-(\ref{scf5}) and (\ref{scf6}), the effective gravitational constant is given by
\begin{equation}
G_{\rm eff} = \frac{1}{16\pi\phi}\left[1+\frac{(1+\xi(\phi)\dot{\phi}^2)^2}{J}\right],
\end{equation}
where
\begin{equation}
J \equiv 3+2\omega+\phi^2\xi(\phi)\left[4\frac{\ddot{\phi}}{\phi}-2\frac{\dot{\phi}^2}{\phi^2}+8H\frac{\dot{\phi}}{\phi}-\phi^2\xi(\phi)\frac{\dot{\phi}^4}{\phi^4}\right].
\end{equation}
The effective gravitational constant $G_{\rm eff}$ is close to Newton's constant $G$ at early times, but increases at later times.

\section{COSMIC GROWTH RATE \label{growthrate}}

\subsection{Density perturbations}

Under the quasistatic approximation on sub-horizon scales, the evolution equation for the cold dark matter overdensity $\delta$ in linear theory is given by
\begin{equation}
\ddot{\delta}+2H\dot{\delta}-4\pi G_{\rm eff}\rho\delta\simeq0,
\end{equation}
where $G_{\rm eff}$ represents the effective gravitational constant of modified gravity models described in the previous section.

We set the initial conditions $\delta\approx a$ and $\dot{\delta}\approx\dot{a}$ at early times. Since we are interested in the difference between the growth of density perturbations in modified gravity and that in the $\Lambda$CDM case, we assume that the initial conditions of matter density perturbations are the same as in the conventional $\Lambda$CDM model. By solving the evolution equation numerically, we obtain the growth factor $\delta/a$ for modified gravity models.
The linear growth rate is written as
\begin{equation}
f=\frac{{d}\ln{\delta}}{{d}\ln{a}}. \label{f_theory}
\end{equation}
where $\delta$ is the matter density fluctuations and $a$ is the scale factor. The growth rate can be parameterized by the growth index $\gamma$, as defined by
\begin{equation}
f=\Omega_m^\gamma.
\end{equation}


Refs. \cite{hir2011b,hirano2012} have shown that the growth rate $f$ in the Galileon model specified by Eqs. (\ref{scf2})-(\ref{scf5}) and (\ref{scf6}) is enhanced compared with the $\Lambda$CDM case, for the same value of $\Omega_{m,0}$, because of the enhancement of effective gravitational constant.

\subsection{Euclid}

Euclid \cite{amendola2012} is a European Space Agency medium class mission, and it is scheduled to be launched in 2019. The main purpose of Euclid is to study the origin of the accelerated expansion of the universe. Euclid will investigate the expansion history and the evolution of cosmic structures by measuring redshifts of galaxies and the distribution of clusters of galaxies over a large portion of the sky. Although its main subject of research is the nature of dark energy, Euclid will cover topics including cosmology, galaxy evolution, and planetary research.

\begin{figure}[h!]
\includegraphics[width=99mm]{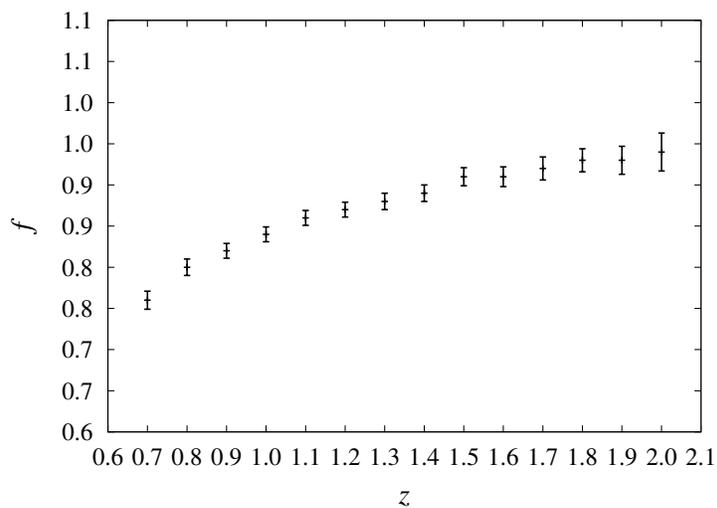}
\caption{Plot of the mock data of the cosmic growth rate.
\label{fig1}}
\end{figure}

In this study, Euclid parameters are adopted as the growth rate observations. The growth rate can be parameterized using the growth index $\gamma$, as defined by $f = {\Omega_m}^\gamma$. Mock data of the cosmic growth rate are created in accordance with the $1\sigma$ marginalized errors of the growth rate that will be used by Euclid, which are shown in Table 4 in Ref. by Amendola et al. \cite{amendola2012}. Table \ref{euclid} lists the $1\sigma$ marginalized errors for the cosmic growth rates in each redshift bin based on Table 4 in the study by Amendola et al. \cite{amendola2012}. In Fig. \ref{fig1}, the mock data of the cosmic growth rate used in this study are plotted.

\begin{table}[h!]
\caption{$1\sigma$ marginalized errors for the growth rates in each redshift bin based on Table 4 in Ref. by Amendola et al. \cite{amendola2012}. Here $z$ represents the redshift and $\sigma_{f_g}$ represents the $1\sigma$ marginalized errors of the growth rates.
\label{euclid}}
\begin{tabular}{c c c c c c}
\hline
\hline
Experimental Parameters & $z$ & $\sigma_{f_g}$(ref.) \\
\hline
Data from Euclid \cite{amendola2012} & 0.7 & 0.011 \\
 & 0.8 & 0.010 \\
 & 0.9 & 0.009 \\
 & 1.0 & 0.009 \\
 & 1.1 & 0.009 \\
 & 1.2 & 0.009 \\
 & 1.3 & 0.010 \\
 & 1.4 & 0.010 \\
 & 1.5 & 0.011 \\
 & 1.6 & 0.012 \\
 & 1.7 & 0.014 \\
 & 1.8 & 0.014 \\
 & 1.9 & 0.017 \\
 & 2.0 & 0.023 \\
\hline
\hline
\end{tabular}
\end{table}

The mock data are used to compute the statistical $\chi^2$ function. $\chi^2$ for the growth rate is defined as
\begin{equation}
\displaystyle \chi_f^2 = \sum_{i=1}^{14}\frac{(f_{theory}(z_i)-f_{obs}(z_i))^2}{\sigma_{f_g}(z_i)^2}
\end{equation}
where $f_{obs}(z_i)$ are the future observational (mock) data of the growth rate. The theoretical growth rate $f_{theory}(z_i)$ is computed as equation (\ref{f_theory}).
In Ref. \cite{hirano2015}, constraints on neutrino masses are estimated based on future observations of the growth rate of cosmic structure from the Euclid redshift survey.

The estimated errors from observational technology of Euclid are known, but the center value of the future observations is not known. Therefore, The purpose of this study is not to know whether $\Lambda$CDM model or modified gravity is valid, but to find ways and probability of distinguishing between $\Lambda$CDM model and modified gravity.

\section{COMPARISON WITH OBSERVATIONS \label{comparison}}

\subsection{extended DGP model}

\begin{figure}[h!]
\includegraphics[width=81mm]{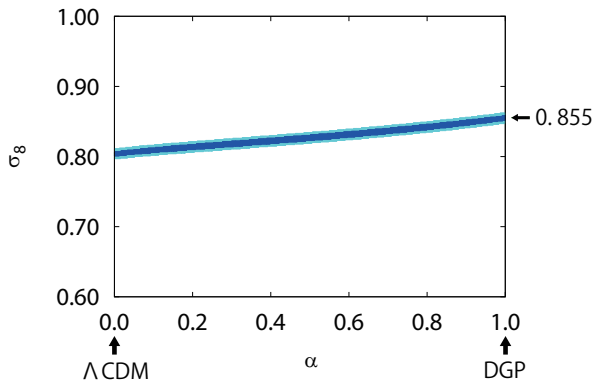}
\caption{Probability contours in the ($\alpha$, $\sigma_8$)-plane for the extended DGP model, from the observational (mock) data of the cosmic growth rate by the Euclid. The contours show the 1$\sigma$ (68.3\%) and 2$\sigma$ (95.0\%) confidence limits.
\label{fig2}}
\end{figure}

\begin{figure}[h!]
\includegraphics[width=81mm]{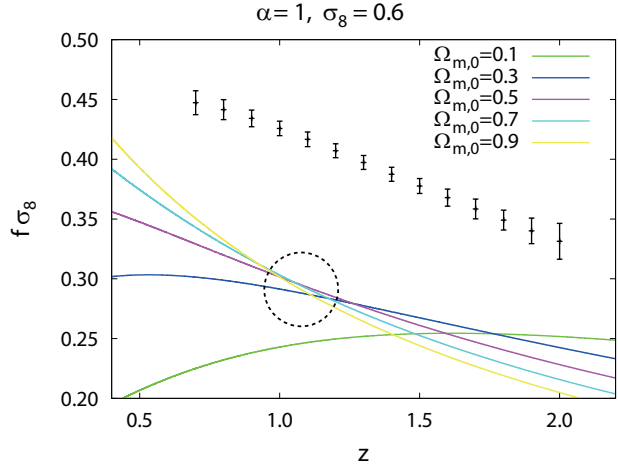}
\caption{$f\sigma_8$ (the product of growth rate and $\sigma_8$) in extended DGP model as a function of redshift $z$ for various values of $\Omega_{m,0}$. the parameters are fixed by $\alpha=1, \sigma_8=0.6$.
\label{fig3}}
\end{figure}

\begin{figure}[h!]
\includegraphics[width=81mm]{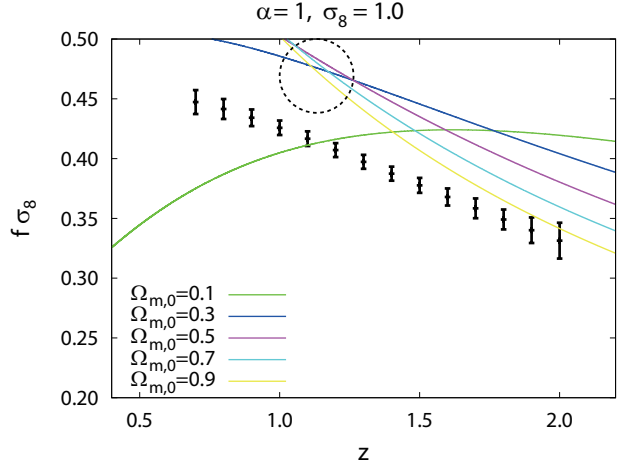}
\caption{$f\sigma_8$ in extended DGP model as a function of redshift $z$ for various values of $\Omega_{m,0}$. the parameters are fixed by $\alpha=1, \sigma_8=1.0$.
\label{fig4}}
\end{figure}

\begin{figure}[h!]
\includegraphics[width=81mm]{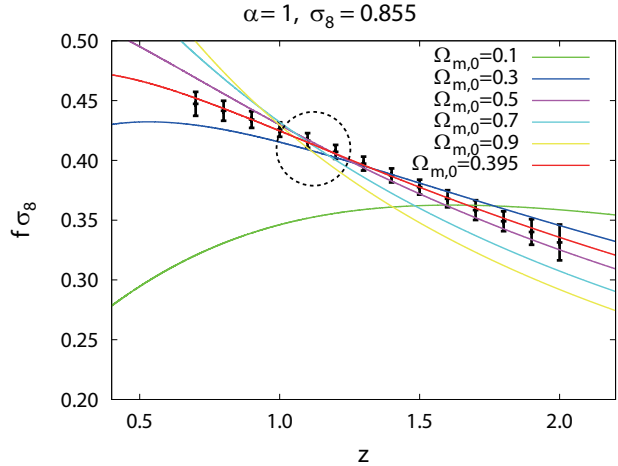}
\caption{$f\sigma_8$ in extended DGP model as a function of redshift $z$ for various values of $\Omega_{m,0}$. the parameters are fixed by $\alpha=1, \sigma_8=0.855$.
\label{fig5}}
\end{figure}

In Fig. \ref{fig2}, we plot the probability contours in the ($\alpha$, $\sigma_8$)-plane in the extended DGP model from the observational (mock) data of the cosmic growth rate by the Euclid.
The blue and light blue contours show the 1$\sigma$ (68.3\%) and 2$\sigma$ (95.0\%) confidence limits, respectively.
Part of $\alpha=0$ for the horizontal axis correspond to the $\Lambda$CDM model, and part of $\alpha=1$ correspond to the original DGP model. $\sigma_8$ is the rms amplitude of over-density at the comoving 8 $h^{-1}$ Mpc scale ($h$ is the normalized Hubble parameter $H_0 = 100 h~{\rm km~sec^{-1}~Mpc^{-1}}$). We demonstrate why $\sigma_8$ is stringently constrained, in Figs. \ref{fig3}, \ref{fig4}, \ref{fig5}.

We plot $f\sigma_8$ (the product of growth rate and $\sigma_8$) in extended DGP model as a function of redshift $z$ for various values of the energy density parameter of the matter at present $\Omega_{m,0}$ in Figs. \ref{fig3}, \ref{fig4}, \ref{fig5}. In Fig. \ref{fig3}, the parameters are fixed by $\alpha=1, \sigma_8=0.6$. For the various values of $\Omega_{m,0}$, theoretical curves seem to revolve around the black-dashed circle. Hence the value of $\sigma_8=0.6$ is incompatible with observational (mock) data.

In Fig. \ref{fig4}, the parameters are fixed by $\alpha=1, \sigma_8=1.0$. For the various values of $\Omega_{m,0}$, theoretical curves seem to revolve around the black-dashed circle. Hence the value of $\sigma_8=1.0$ is incompatible with observational (mock) data.

In Fig. \ref{fig5}, the parameters are fixed by $\alpha=1, \sigma_8=0.855$. For the various values of $\Omega_{m,0}$, although theoretical curves seem to revolve around the black-dashed circle, there are theoretical curves comparatively close to observational (mock) data. Hence the value of $\sigma_8=0.855$ is compatible with observational (mock) data in original DGP model ($\alpha=1$).

\begin{figure}[h!]
\includegraphics[width=81mm]{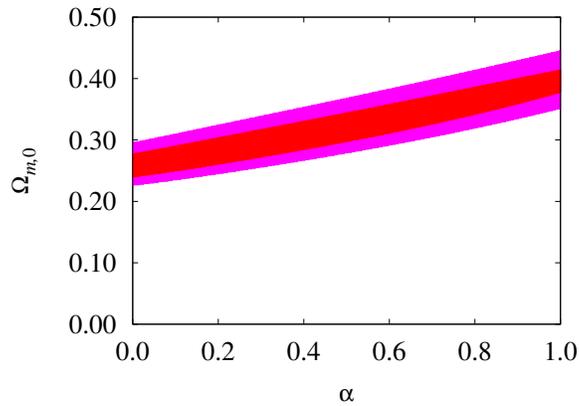}
\caption{Probability contours in the ($\alpha$, $\Omega_{m,0}$)-plane for the extended DGP model, from the observational (mock) data of the cosmic growth rate by the Euclid. The contours show the 1$\sigma$ (68.3\%) and 2$\sigma$ (95.0\%) confidence limits.
\label{fig6}}
\end{figure}

\begin{figure}[h!]
\includegraphics[width=81mm]{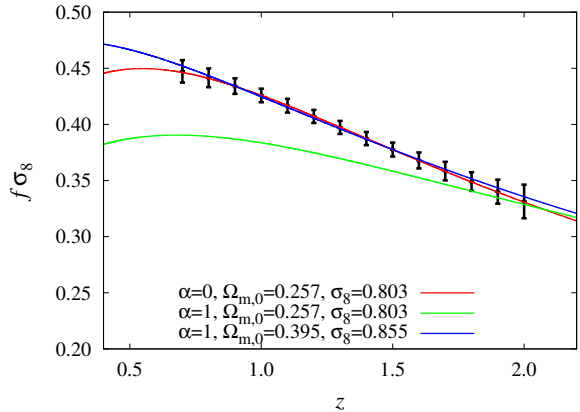}
\caption{$f\sigma_8$ in extended DGP model as a function of redshift $z$. The values of the parameters are as follows. Red line: $\alpha$=0, $\Omega_{m,0}$=0.257, $\sigma_8$=0.803. Green line: $\alpha$=1, $\Omega_{m,0}$=0.257, $\sigma_8$=0.803. Blue line: $\alpha$=1, $\Omega_{m,0}$=0.395, $\sigma_8$=0.855. 
\label{fig7}}
\end{figure}

In Fig. \ref{fig6}, we plot the probability contours in the ($\alpha$, $\Omega_{m,0}$)-plane in the extended DGP model from the observational (mock) data of the cosmic growth rate by the Euclid.
The red and pink contours show the 1$\sigma$ (68.3\%) and 2$\sigma$ (95.0\%) confidence limits, respectively.
We demonstrate why $\Omega_{m,0}$ be positively correlated with $\alpha$, in Fig. \ref{fig7}.

We plot $f\sigma_8$ in extended DGP model as a function of redshift $z$ in Fig. \ref{fig7}. Red line is theoretical curve for best fit parameter in $\Lambda$CDM model ($\alpha$=0, $\Omega_{m,0}$=0.257, $\sigma_8$=0.803). In case of changing only to $\alpha$=1, the growth rate $f\sigma_8$ is suppressed, because of the suppression of effective gravitational constant (Green line: $\alpha$=1, $\Omega_{m,0}$=0.257, $\sigma_8$=0.803). For $\alpha$=1, tuning the value of $\Omega_{m,0}$ and $\sigma_8$, theoretical curve is compatible with observational (mock) data again (Blue line: $\alpha$=1, $\Omega_{m,0}$=0.395, $\sigma_8$=0.855).

\begin{figure}[h!]
\includegraphics[width=81mm]{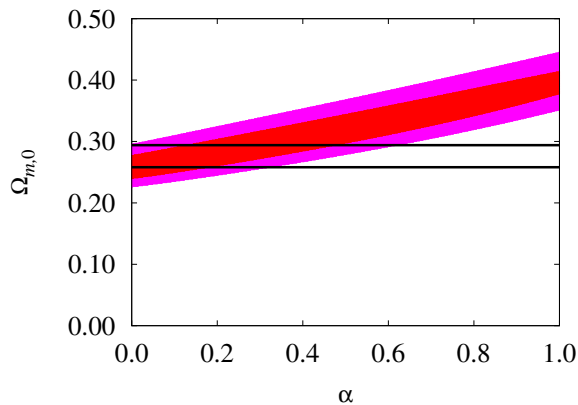}
\caption{Addition constraints on $\Omega_{m,0}$ for the extended DGP model form CMB, BAO and SNIa data \cite{xia2009} to Fig. \ref{fig6}.
\label{fig8}}
\end{figure}

In Fig. \ref{fig8}, we add constraints on $\Omega_{m,0}$ for the extended DGP model form the combination of CMB, BAO and SNIa (black lines) \cite{xia2009} to probability contours in the ($\alpha$, $\Omega_{m,0}$)-plane by growth rate (mock) data by the Euclid of Fig. \ref{fig6}. 

Since $\Omega_{m,0}$ is stringently constrained by the cosmic growth rate data from the Euclid, we find the $\Lambda$CDM model is distinguishable from original DGP model by combining the growth rate data of the Euclid with other observations.

\subsection{Kinetic Gravity Braiding model}

\begin{figure}[h!]
\includegraphics[width=81mm]{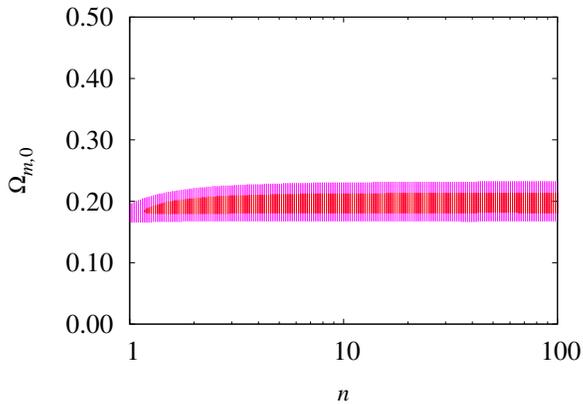}
\caption{Probability contours in the ($n$, $\Omega_{m,0}$)-plane for the kinetic gravity braiding model, from the observational (mock) data of the cosmic growth rate by the Euclid. The contours show the 1$\sigma$ (68.3\%) and 2$\sigma$ (95.0\%) confidence limits.
\label{fig9}}
\end{figure}

In Fig. \ref{fig9}, we plot the probability contours in the ($n$, $\Omega_{m,0}$)-plane in the kinetic gravity braiding model from the observational (mock) data of the cosmic growth rate by the Euclid. 
The red and pink contours show the 1$\sigma$ (68.3\%) and 2$\sigma$ (95.0\%) confidence limits, respectively.

\begin{figure}[h!]
\includegraphics[width=81mm]{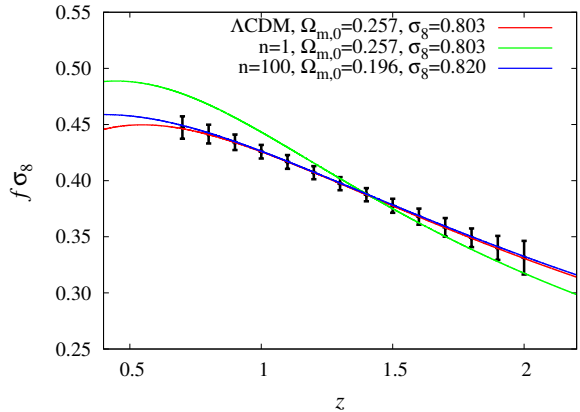}
\caption{$f\sigma_8$ in kinetic gravity braiding model as a function of redshift $z$. The values of the parameters are as follows. Red line: $\Lambda$CDM, $\Omega_{m,0}$=0.257, $\sigma_8$=0.803. Green line: $n$=1, $\Omega_{m,0}$=0.257, $\sigma_8$=0.803. Blue line: $n$=100, $\Omega_{m,0}$=0.196, $\sigma_8$=0.820.
\label{fig10}}
\end{figure}

We plot $f\sigma_8$ in kinetic gravity braiding model as a function of redshift $z$ in Fig. \ref{fig10}. Red line is theoretical curve for best fit parameter in $\Lambda$CDM model ($\Omega_{m,0}$=0.257, $\sigma_8$=0.803). In case of kinetic gravity braiding model for $n$=1, the growth rate $f\sigma_8$ is enhanced because of the enhancement of effective gravitational constant (Green line: $n$=1, $\Omega_{m,0}$=0.257, $\sigma_8$=0.803). For $n$=100, tuning the value of $\Omega_{m,0}$ and $\sigma_8$, theoretical curve is compatible with observational (mock) data again (Blue line: $n$=100, $\Omega_{m,0}$=0.196, $\sigma_8$=0.820).

\begin{figure}[h!]
\includegraphics[width=81mm]{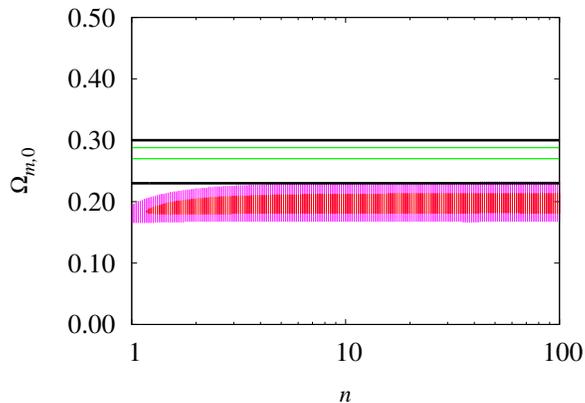}
\caption{Adding constraints on $\Omega_{m,0}$ for the kinetic gravity braiding model form CMB (green lines) and from SNIa (black lines) respectively \cite{kim2011} to Fig. \ref{fig9}.
\label{fig11}}
\end{figure}

In Fig. \ref{fig11}, we add constraints on $\Omega_{m,0}$ for the kinetic gravity braiding model form CMB (green lines) and from SNIa (black lines) respectively \cite{kim2011} to probability contours in the ($n$, $\Omega_{m,0}$)-plane by growth rate (mock) data by the Euclid of Fig. \ref{fig9}.

In the kinetic gravity braiding model, the allowed parameter region obtained using only the growth rate data does not overlap with the allowed parameter region obtained from CMB, or from SNIa data.

\subsection{Galileon model}

\begin{figure}[h!]
\includegraphics[width=81mm]{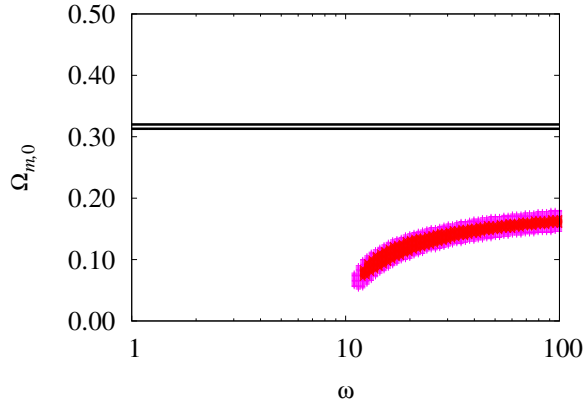}
\caption{Probability contours in the ($\omega$, $\Omega_{m,0}$)-plane for the Galileon model, from the observational (mock) data of the cosmic growth rate by the Euclid. The contours show the 1$\sigma$ (68.3\%) and 2$\sigma$ (95.0\%) confidence limits. Adding constraints on $\Omega_{m,0}$ for the Galileon model form the combination of CMB, BAO and SNIa (black lines) \cite{hirano2012}.
\label{fig12}}
\end{figure}

In Fig. \ref{fig12}, we plot the probability contours in the ($\omega$, $\Omega_{m,0}$)-plane in the Galileon model from the observational (mock) data of the cosmic growth rate by the Euclid. 
The red and pink contours show the 1$\sigma$ (68.3\%) and 2$\sigma$ (95.0\%) confidence limits, respectively.
In addition, we plot constraints on $\Omega_{m,0}$ for the Galileon model form the combination of CMB, BAO and SNIa (black lines) \cite{hirano2012}.

In the Galileon model, the allowed parameter region obtained using only the growth rate data does not overlap with the allowed parameter region obtained from the combination of CMB, BAO, and SNIa data at all.

\section{CONCLUSIONS \label{conclusion}}
The growth rate $f\sigma_8$ in the original DGP model is suppressed compared with the $\Lambda$CDM case, for the same value of $\Omega_{m,0}$, because of the suppression of effective gravitational constant.
In case of the kinetic gravity braiding model and the Galileon model, the growth rate $f\sigma_8$ is enhanced compared with the $\Lambda$CDM case, for the same value of $\Omega_{m,0}$, because of the enhancement of effective gravitational constant.
For future observational data of the cosmic growth rate, compatible value of $\Omega_{m,0}$ are different by models, furthermore value of $\Omega_{m,0}$ can be stringently constrained. Thus, we find the $\Lambda$CDM model is distinguishable from modified gravity by combining the growth rate data of the Euclid with other observations.

The estimated errors from observational technology of Euclid are known, but the center value of the future observations is not known. 
If the center value of the cosmic growth rate the future observations is different from that of this paper, valid model can be different from that of this paper. Although the way of this paper is useful in older to distinguishing between $\Lambda$CDM model and modified gravity, and the $\Lambda$CDM model is distinguishable from modified gravity. 

In this paper, Assuming the function $G_5(\phi,X)$ in the Horndeski's theory $G_5(\phi,X)=0$, and we compute liner matter density perturbations for the growth rate.
In future work, we study the model having the function $G_5(\phi,X)$ in the Horndeski's theory, and we investigate the nonliner effect.   

\bibliography{koichi}

\end{document}